\documentclass[letterpaper,journal]{IEEEtran}
\usepackage{amsmath,amsfonts}
\usepackage{algorithmic}
\usepackage{algorithm}
\usepackage{array}
\usepackage[caption=false,font=normalsize,labelfont=sf,textfont=sf]{subfig}
\usepackage{textcomp}
\usepackage{stfloats}
\usepackage{multirow}
\usepackage{url}
\usepackage{verbatim}
\usepackage{graphicx}
\usepackage{cite}
\usepackage{booktabs}
\usepackage[utf8]{inputenc}
\usepackage{microtype}
\usepackage{pifont}
\usepackage{makecell}

\begin{document}

\title{Articulatory Source-Filter TTS: Physically Grounded Control through Vocal Tract Kinematics}

\author{Jesuraj~Bandekar,
        Shinji~Watanabe,~\IEEEmembership{Fellow,~IEEE,}
        and~Prasanta~Kumar~Ghosh,~\IEEEmembership{Senior Member,~IEEE}%
\thanks{Jesuraj Bandekar and Prasanta Kumar Ghosh are with the SPIRE Lab, Department of Electrical Engineering, Indian Institute of Science (IISc), Bangalore 560012, India (e-mail: jesurajbandekar@gmail.com; prasantg@iisc.ac.in).}
\thanks{Shinji Watanabe is with the Language Technologies Institute, Carnegie Mellon University, Pittsburgh, PA 15213, USA (e-mail: shinjiw@ieee.org).}%

}




\markboth{IEEE/ACM Transactions on Audio, Speech, and Language Processing}%
{Bandekar \MakeLowercase{\textit{et al.}}: Articulatory Source-Filter TTS}

\maketitle

\begin{abstract}
Modern neural text-to-speech (TTS) systems achieve remarkable acoustic fidelity but operate as opaque black boxes, offering limited interpretable control over the physical speech production process. While some models allow prosodic manipulation, explicit control over the vocal tract filter remains largely inaccessible. In this paper, we propose a novel, controllable source-filter TTS architecture that grounds the synthesis process in physical articulatory kinematics. Our approach utilises an Acoustic-to-Articulatory Inversion (AAI) model, enhanced by large-scale pretrained representations, to generate kinematic pseudo-trajectories for a large TTS corpus. These trajectories explicitly condition the filter response, while predicted pitch and energy contours parameterise the glottal source. The source and filter are predicted independently, and the source is refined by an Optimal Transport Conditional Flow Matching (OT-CFM) module before recombination with the filter to form the final spectrogram. Experimental results demonstrate that our model achieves intelligibility and naturalness competitive with similarly sized baselines, incurring only a modest spectral fidelity cost from constraining the output to an explicit source-filter decomposition, in exchange for control unavailable to black-box systems. Furthermore, quantitative and qualitative evaluations reveal explicit source-filter disentanglement, allowing for stable prosodic scaling and cross-speaker source/filter recombination. In this recombination, prosodic source attributes, notably F0, remain tied to the source speaker while vocal-tract spectral characteristics consistently follow the filter speaker. Finally, we demonstrate that the model enables explicit, fine-grained articulatory control, such as accent modification through direct spatial manipulation of articulatory trajectories, offering a new direction for interpretable speech synthesis. Audio samples are available at \url{https://coding-phoenix-12.github.io/ArticulatorySFTTS/}.
\end{abstract}


\section{Introduction}
\IEEEPARstart{M}{odern} text-to-speech (TTS) systems generate audio of very high naturalness and acoustic quality. Recent methods based on flow-matching, neural audio codecs, and Large Language Models (LLMs) \cite{chen2025f5, wang2025maskgct, liu2025e1} use large architectures and vast training data to produce zero-shot synthesis that is nearly indistinguishable from natural human speech. This progress, however, comes at the cost of interpretability, as these systems function largely as black boxes. Despite their strong performance, they offer little control or insight for downstream tasks such as dialect understanding, accent modification, voice conversion, and clinical and pathological speech modelling, all of which require fine-grained, disentangled control. Such applications benefit from physically interpretable control: an articulatory handle on the vocal tract allows dialectal variation to be edited in gesture space, vocal-tract identity to be separated from prosody, and atypical articulation to be probed in physical terms, none of which is accessible in a black-box latent space. 


As summarised in Table~\ref{tab:intro_comparison}, recent controllable models expose prosodic attributes such as pitch and energy~\cite{li2023styletts, lee2025hierspeech++, ju2024naturalspeech}, and source--filter approaches go further by separating source and filter spectrograms~\cite{bak2021fastpitchformant, han2025improving}, yet none provides an explicit, physically grounded handle on the vocal tract response, nor any connection to well-established speech production theory. In this work, we argue that recovering physically grounded control while preserving high-quality synthesis requires a glottal-source and articulation-based decomposition, which would in turn enable explicit physical control.

The source–filter model of speech production \cite{fant1971acoustic} provides the theoretical foundation for this decomposition, separating speech into a glottal source that carries pitch and energy, and a vocal-tract filter that shapes the final spectral output. Early formant synthesisers \cite{klatt1980software} modelled both explicitly, yielding fully interpretable but mechanical speech, and more recent physical synthesisers such as VocalTractLab~\cite{birkholz2013modeling, gao2024articulatory} improve quality but remain driven by hand-designed gestures or recorded audio rather than text. Subsequent neural approaches \cite{hunt1996unit, zen2009statistical, wang2017tacotron, ren2019fastspeech, kim2020glow, kim2021conditional, wang2023neural} abandoned this physical grounding in favour of acoustic fidelity, mapping text directly to acoustics through opaque latent representations (Table~\ref{tab:intro_comparison}).


While some systems expose pitch and energy to control the glottal source (Table~\ref{tab:intro_comparison}), the vocal tract filter lacks a comparable physical parameterisation in TTS. Real-time MRI (rtMRI) \cite{narayanan2004approach, narayanan2014real} captures full articulatory data for synthesis \cite{toutios2016articulatory, otani2023speech}, but its high cost and acoustic reverberation make it a poor training target, whereas Electromagnetic Articulography (EMA) offers a tractable alternative. Because EMA corpora are small, extensive work has leveraged Self-Supervised Learning (SSL) representations and auxiliary-task pretraining to build robust Acoustic-to-Articulatory Inversion (AAI) models \cite{illa2018low, udupa2021estimating, hao2024exploring, bandekar2025enhancing}, though their pseudo-targets are mostly applied in speech-to-speech resynthesis or coding frameworks \cite{cho2024coding, liu2025rt}. We instead integrate articulatory trajectories into a TTS architecture, explicitly grounding the filter of a source-filter decomposition in the physical kinematics of the vocal tract.

Building on this, we propose a controllable source-filter TTS architecture whose filter is grounded in AAI-supervised articulatory trajectories, with the following contributions:
\begin{enumerate}
    \item \textbf{A controllable source--filter TTS architecture} in which the filter is conditioned on predicted articulatory trajectories, providing, to our knowledge, the first text-driven TTS with physically grounded, kinematic-level control over the vocal-tract response.
    \item \textbf{A two-stage AAI-based pseudo-labelling pipeline} using multilingual speech-model features that scales articulatory supervision from a small EMA corpus to hundreds of hours of unpaired TTS speech.
    \item \textbf{Competitive synthesis quality}, showing that the explicit source--filter decomposition retains intelligibility and naturalness competitive with comparably-sized baselines at only a modest spectral-fidelity cost.
    \item \textbf{Quantitative source--filter disentanglement} through (a) pitch/energy scaling, where filter-side properties (formants, spectral envelope) remain stable under source perturbation, and (b) cross-speaker swapping, where pitch stays tied to the source speaker while spectral characteristics follow the filter speaker.

    \item \textbf{Qualitative physical interpretability} through articulatory--acoustic analyses, including articulator decoupling across minimal pairs and recovery of the canonical American rhotic posture in post-vocalic /r/.
    \item \textbf{Interpretable accent modification}, converting a rhotic American /r/ to its non-rhotic realisation by flattening the tongue-tip and tongue-body gestures, with the corresponding $F3$ shift in the synthesised audio control unavailable in black-box neural TTS.
\end{enumerate}

The remainder of the paper covers related work (Section~\ref{sec:Background}), the proposed methodology including the AAI backend and TTS architecture (Section~\ref{sec:method}), the experimental setup (Section~\ref{sec:exp_setup}), the evaluation framework (Section~\ref{sec:metrics}), and results with analysis (Section~\ref{sec:results}).

\begin{table}[htb]
\centering
\caption{Comparison of representative speech synthesis approaches by input modality, exposed interpretable representations, and controllable aspects. Only the proposed model is text-driven with an explicit source/filter decomposition and articulatory filter control. Legend below the table.}
\label{tab:intro_comparison}
\renewcommand{\arraystretch}{1.05}
\setlength{\tabcolsep}{4pt}
\begin{tabular}{@{}llcc@{}}
\toprule
\textbf{Model} & \textbf{Input} & \textbf{Interp.} & \textbf{Control} \\
\midrule
\multicolumn{4}{c}{\textit{Black-box neural TTS}} \\
Tacotron~\cite{wang2017tacotron}      & Text & --        & --      \\
FastSpeech~\cite{ren2019fastspeech}   & Text & --        & --      \\
Glow-TTS~\cite{kim2020glow}           & Text & --        & --      \\
VITS~\cite{kim2021conditional}        & Text & --        & --      \\
VALL-E~\cite{wang2023neural}          & Text & --        & --      \\
StyleTTS2~\cite{li2023styletts}       & Text & P, E        & P, E       \\
HierSpeech++~\cite{lee2025hierspeech++} & Text & P      & P       \\
NaturalSpeech~3~\cite{ju2024naturalspeech} & Text & AC & P, T \\
\midrule
\multicolumn{4}{c}{\textit{Source--filter TTS}} \\
FastPitch~\cite{lancucki2021fastpitch}    & Text & P, E     & P, E    \\
FastSpeech~2~\cite{renfastspeech}         & Text & P, E     & P, E    \\
FastPitchFormant~\cite{bak2021fastpitchformant} & Text & SF, P & P \\
StableForm-TTS~\cite{han2025improving}    & Text & SF, P, E     & P, E    \\
\midrule
\multicolumn{4}{c}{\textit{Physical / articulatory synthesis}} \\
VocalTractLab~\cite{birkholz2013modeling} & Gestures & VT & AG \\
ACS~\cite{gao2024articulatory}            & Audio    & VT & AG \\
\midrule
\textbf{Proposed} & \textbf{Text} & \textbf{P, E, AT, SF} & \textbf{P, E, AT} \\
\bottomrule
\end{tabular}
\vspace{1mm}
\\ \footnotesize\raggedright
P: pitch; E: energy; AT: articulatory trajectories; SF: source/filter spectrograms; VT: vocal-tract geometry/parameters; AG: articulatory gestures; AC: attribute codes; T: timbre.
\vspace{-2mm}
\end{table}

\begin{figure*}[htbp]
    \centering
    \includegraphics[width=\textwidth]{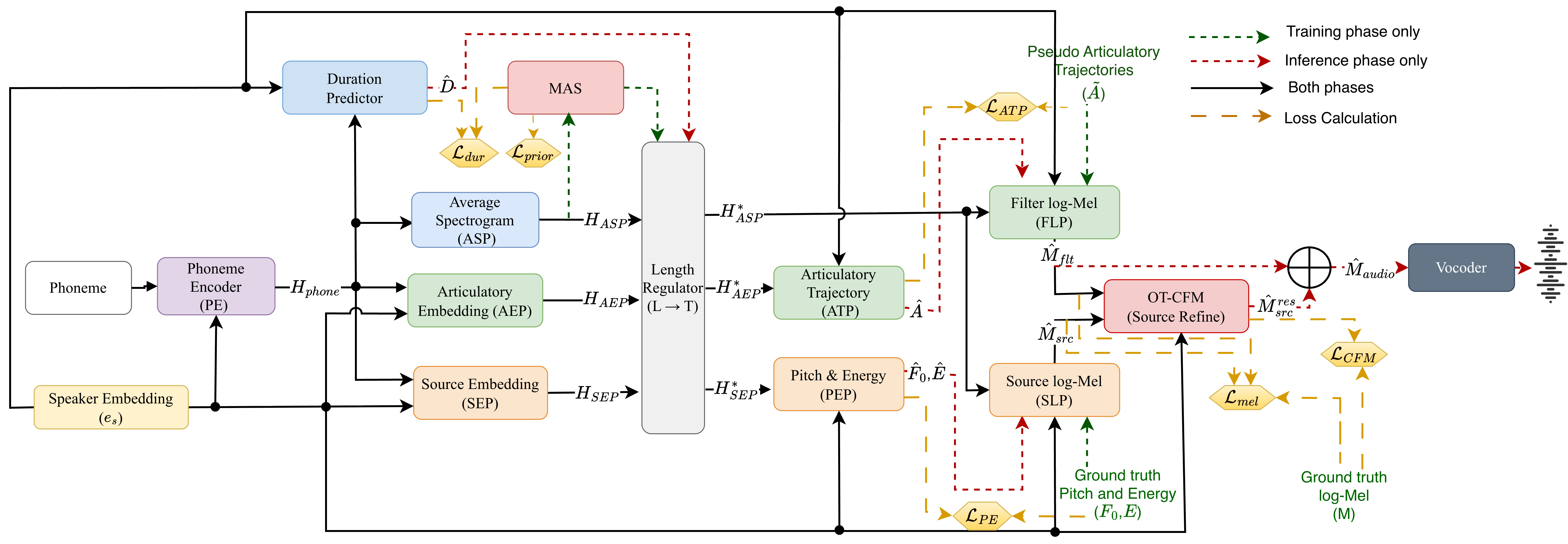}
    \vspace{-7mm} 
    \caption{Overview of the proposed controllable source-filter TTS architecture. Solid lines denote data flow active during both training and inference. Green dashed lines indicate paths exclusive to the training phase (e.g., MAS alignment, AAI pseudo-target supervision), while red dashed lines represent the inference-only generative paths. Optimisation pathways for the multi-task objective are highlighted in yellow.}
    \label{fig:architecture}
      \vspace{-5mm}
\end{figure*}

\section{Background and Related Work}
\label{sec:Background}

\subsection{Flow Matching based TTS}

Flow Matching (FM) considers a prior noise distribution $x_0 \sim q_0$ (typically a standard Gaussian) and a target data distribution $x_1 \sim p_{\text{data}}$, learning a time-dependent neural vector field that transports samples from $q_0$ to $p_{\text{data}}$ over $\tau \in [0,1]$. Since marginalising over the full data distribution is intractable, Conditional Flow Matching (CFM) conditions the vector field on individual samples $x_1$. Optimal Transport CFM (OT-CFM) further constrains the conditional path between noise and data to a deterministic straight line, $\phi_\tau(x) = (1-\tau)x_0 + \tau x_1$, yielding a constant conditional vector field $u_\tau(x \mid x_1) = x_1 - x_0$ (see \cite{lipman2023flow} for the full derivation). For conditional generative tasks such as TTS, the neural vector field is further conditioned on auxiliary variables $c$ (e.g., linguistic features, prosodic cues, or articulatory representations). The resulting training objective is
\begin{equation}
\mathcal{L}_{\text{OT-CFM}}(\theta) = \mathbb{E}_{\tau, x_0, x_1} \!\left[\, \|(x_1 - x_0) - v_\tau(\phi_\tau(x); \theta, c)\|^2 \,\right].
\end{equation}
where $v_\tau$ is the neural vector field and $\theta$ the learnable parameters.

Models such as Matcha-TTS \cite{mehta2024matcha}, Voicebox \cite{le2023voicebox}, and P-Flow \cite{kim2023p} apply this framework to TTS, mapping prior noise to acoustic features (typically Mel-spectrograms) conditioned on linguistic encodings, durations, and speaker embeddings, achieving strong zero-shot naturalness with simulation-free training.

\subsection{Controllable and Interpretable Source-Filter TTS}

Beyond black-box mapping, several architectures expose source-filter attributes for interpretable control. Early non-autoregressive models such as FastPitch \cite{lancucki2021fastpitch} and FastSpeech~2 \cite{renfastspeech} introduced explicit pitch and energy predictors. More recent systems \cite{li2025styletts, lee2025hierspeech++, ellinas2023controllable} combine these with style or prosodic embeddings for finer control. A complementary direction generates source and filter spectrograms separately and combines them. FastPitchFormant \cite{bak2021fastpitchformant} couples a pitch-driven excitation generator with a formant generator, and \cite{han2025improving} extends this with a diffusion-based source decoder and energy prediction. While these approaches provide interpretable control over source attributes, they treat the filter as an abstract spectral object with no physical correspondence to the articulators.

\subsection{Acoustic-to-Articulatory Inversion and Articulatory Speech Synthesis}
Acoustic-to-articulatory inversion (AAI) is the task of estimating the physical movements of the vocal tract articulators from a recorded speech signal. This data of the articulator movement is collected using an Electromagnetic Articulograph (EMA). While AAI solves the inverse problem of uncovering physical states from acoustics, Articulatory Speech Synthesis or forward mapping, focuses on the generation of high-fidelity speech directly from these articulatory trajectories.
Both these allow for a comprehensive modelling of the relationship between vocal tract biomechanics and the resulting acoustic output.

Deep learning-based methods for AAI include Convolutional Neural Networks (CNN), Recurrent Neural Networks (RNN) based sequence-to-sequence modeling \cite{illa2018low, bozorg2020acoustic, illa2019representation}, and Transformer-based networks \cite{udupa2021estimating, udupa2022streaming, chung2024speaker}. These models use audio representations, such as spectrograms and Mel-frequency cepstral coefficients (MFCCs), to predict EMA trajectories. However, the scarcity of EMA data poses a considerable bottleneck for achieving high-level performance, especially for unseen speakers. To tackle this, recent methods such as utilising self-supervised learning (SSL) representations instead of standard audio features \cite{udupa2023improved, maharana2024acoustic, hao2024exploring, cho2024self, cho2023evidence} and employing auxiliary task pretraining \cite{bandekar2025enhancing} have been shown to significantly improve performance. Notably, SSL representations have been shown to implicitly encode articulatory information \cite{cho2024self, cho2023evidence}, which contributes to their strong AAI performance.

Similar models have been explored for articulatory speech synthesis. \cite{bandekar2024articulatory, chen2021ema2s, wu2022deep} use phoneme-based losses, multimodal training, and GAN-based architectures, while \cite{liu2024fast} achieves highly efficient synthesis using differentiable digital signal processing (DDSP). Additionally, \cite{cho2024coding} and \cite{liu2025rt} use articulatory trajectories as intermediate controls for pronunciation modification and voice conversion, respectively.

\section{Proposed Methodology}
\label{sec:method}

 We use a two-stage setup. First, an AAI model is trained on parallel EMA-speech data to map acoustics to the physical kinematics of the vocal-tract articulators (the lips, jaw, and tongue), later detailed as 12-channel midsagittal trajectories in Section~\ref{subsubsec:AAI_dataset}. Second, it generates articulatory pseudo-trajectories over a large TTS corpus, which then serve as ground truth to train our source-filter explicit-control TTS model.

\subsection{Articulatory Pseudo-Target Generation via AAI}
\label{subsec:AAI}
For the AAI model, we adopt the Transformer-based architecture of \cite{udupa2021estimating}. Recent literature has extensively demonstrated that utilising representations from large-scale pretrained models significantly enhances AAI performance. We extract acoustic features using the massively multilingual MMS-1B model \cite{pratap2024scaling}. Because this model is pre-trained on over a thousand languages, it encapsulates vast phonetic variability, which has been shown to substantially improve inversion performance, particularly in cross-lingual scenarios \cite{hao2024exploring}.

Furthermore, acoustic and phonetic properties are distributed heterogeneously across the depth of large pre-trained speech models \cite{pasad2021layer}. Therefore, rather than relying on a single hidden state, we extract representations from three distinct depths of the MMS-1B model, specifically layers 6, 24, and 36. These extracted states are concatenated along the feature dimension to form the final composite acoustic input. Given a 16 kHz audio input, the MMS-1B model yields feature frames at a rate of 50 Hz.

Given an input acoustic feature sequence, $\mathbf{M} = [\mathbf{m}_1, \dots, \mathbf{m}_T]$ with $\mathbf{m}_t \in \mathbb{R}^{D}$ the $D$-dimensional composite MMS feature, and the corresponding ground-truth articulatory trajectories $\mathbf{A} = [\mathbf{a}_1, \dots, \mathbf{a}_T]$, with $\mathbf{a}_t \in \mathbb{R}^{K}$ the $K$-dimensional articulatory frame, both of frame length $T$ (so that $\mathbf{M} \in \mathbb{R}^{D \times T}$ and $\mathbf{A} \in \mathbb{R}^{K \times T}$), the estimated kinematic sequence is predicted as


\begin{equation}
    \tilde{\mathbf{A}} = \mathcal{F}_{\text{AAI}}(\mathbf{M})
    \label{eq:aai_prediction}
\end{equation}

where $\mathcal{F}_{\text{AAI}}$ denotes the AAI model, and $\tilde{\mathbf{A}} = [\tilde{\mathbf{a}}_1, \dots, \tilde{\mathbf{a}}_T]$ represents the generated pseudo-trajectories. These pseudo-trajectories subsequently serve as the supervisory training targets for the Articulatory Trajectory Predictor (ATP), which we detail in \ref{subsec:methodology-tts}.

The model $\mathcal{F}_{\text{AAI}}$ is optimised using two losses. First, the absolute positional error is measured using the Mean Squared Error (MSE) loss, averaged over all $T$ frames and $K$ channels:

\begin{equation}
\mathcal{L}_{\text{MSE}} =
\frac{1}{TK}
\sum_{t=1}^{T}
\sum_{k=1}^{K}
\left(
a_{t,k} - \tilde{a}_{t,k}
\right)^2
\label{eq:aai_mse}
\end{equation}

where $a_{t,k}$ and $\tilde{a}_{t,k}$ represent the ground-truth and predicted values of the $k$-th articulator at time step $t$, respectively.

Second, to capture the shape and dynamic movement of the articulators over time, a Pearson Correlation Coefficient (PCC) loss is used. For two length-$T$ sequences $\mathbf{u},\mathbf{v}\in\mathbb{R}^T$ with temporal means $\bar{u},\bar{v}$, the Pearson correlation is
\begin{equation}
\rho(\mathbf{u}, \mathbf{v}) =
\frac{\sum_{t=1}^{T}(u_t - \bar{u})(v_t - \bar{v})}
{\sqrt{\sum_{t=1}^{T}(u_t - \bar{u})^2 \; \sum_{t=1}^{T}(v_t - \bar{v})^2}}.
\label{eq:pcc_def}
\end{equation}
The PCC loss averages $1-\rho$ over the $K$ articulator channels:
\begin{equation}
\mathcal{L}_{\text{PCC}} =
\frac{1}{K}\sum_{k=1}^{K}\bigl(1 - \rho(\mathbf{a}_k, \tilde{\mathbf{a}}_k)\bigr),
\label{eq:pcc}
\end{equation}
where $\mathbf{a}_k,\tilde{\mathbf{a}}_k\in\mathbb{R}^T$ are the ground-truth and predicted trajectories of channel $k$.

The final training objective for the AAI model is computed as the direct sum of the two penalty terms:

\begin{equation}
\mathcal{L}_{\text{AAI}} =
\mathcal{L}_{\text{MSE}}
+
\mathcal{L}_{\text{PCC}}
\end{equation}

Once training is done, we use this model to generate pseudo-trajectories for the target TTS dataset.

\subsection{Controllable Source-Filter TTS Architecture}
\label{subsec:methodology-tts}

To realise the explicit source-filter-based disentanglement of the speech signal, our core Text-to-Speech (TTS) architecture builds upon the structural foundations of the FastPitchFormant framework \cite{bak2021fastpitchformant} and \cite{han2025improving}. Throughout, $\hat{(\cdot)}$ denotes a prediction, un-hatted symbols denote ground truth, $\tilde{(\cdot)}$ an AAI pseudo-target from Section \ref{subsec:AAI}, and $*$ a length-regulated sequence upsampled to $T$. $\mathbf{M}^{\text{res}}_{\text{src}} = \mathbf{M} - \hat{\mathbf{M}}_{\text{flt}}$ is the residual source target for OT-CFM training which is employed in Section \ref{subsubsec:src_refin}. 

The model relies on a pre-trained speaker embedding module that extracts a global speaker vector, $\mathbf{e}_s \in \mathbb{R}^{d_{\text{spk}}}$, per utterance where $d_{\text{spk}}$ denotes the dimension of the speaker embedding vector. This embedding is explicitly injected into every subsequent block to provide consistent speaker~(Fig. \ref{fig:architecture}). The speaker embedding is injected into every block rather than only at the input so that speaker identity conditions each stage consistently. This is essential for the cross-speaker routing experiments, where source and filter blocks must be driven by different speaker embeddings independently

\subsubsection{Linguistic Encoding and Length Regulation}
\label{subsubsec:L_encode_LR}
Given an input phoneme sequence $\mathbf{P} = [p_1, \dots, p_L]$ of length $L$, where each $p_i$ denotes a phoneme token, the Phoneme Encoder (PE)~(Fig.~\ref{fig:architecture}), parameterised by learnable weights $\theta_{\text{PE}}$, predicts the linguistic hidden representation sequence, $\mathbf{H}_{\text{phone}} \in \mathbb{R}^{d_{\text{hidden}} \times L}$, as:


\begin{equation}
    \mathbf{H}_{\text{phone}} = \mathcal{F}_{\theta_{\text{PE}}}(\mathbf{P}, \mathbf{e}_s)
    \label{eq:h_phone}
\end{equation}

where $d_{\text{hidden}}$ is the dimension of the encoder's hidden state. This phoneme representation is subsequently projected into three distinct intermediate representation streams: the Articulatory Embedding Predictor (AEP), the Source Embedding Predictor (SEP), and the Average Spectrogram Projection (ASP)~(all shown in Fig.~\ref{fig:architecture}), parameterised by learnable weights $\theta_{\text{AEP}}$, $\theta_{\text{SEP}}$, and $\theta_{\text{ASP}}$,  resulting in the sequences $\mathbf{H}_{\text{AEP}}, \mathbf{H}_{\text{SEP}} \in \mathbb{R}^{d_{\text{hidden}} \times L}$, and $\mathbf{H}_{\text{ASP}} \in \mathbb{R}^{d_{\text{mel}} \times L}$, respectively:


\vspace{-0.5cm}
\begin{equation}
    \mathbf{H}_{\text{AEP}} = \mathcal{F}_{\theta_{\text{AEP}}}(\mathbf{H}_{\text{phone}}, \mathbf{e}_s)
    \label{eq:h_aep}
\end{equation}
\vspace{-0.5cm}
\begin{equation}
    \mathbf{H}_{\text{SEP}} = \mathcal{F}_{\theta_{\text{SEP}}}(\mathbf{H}_{\text{phone}}, \mathbf{e}_s)
    \label{eq:h_sep}
\end{equation}
\vspace{-0.5cm}
\begin{equation}
    \mathbf{H}_{\text{ASP}} = \mathcal{F}_{\theta_{\text{ASP}}}(\mathbf{H}_{\text{phone}}, \mathbf{e}_s)
    \label{eq:h_asp}
\end{equation}

We project into three separate streams rather than a shared representation so that the source and filter can be predicted independently for the explicit disentanglement. This branch structure follows the decomposed design of FastPitchFormant \cite{bak2021fastpitchformant}. To resolve the length mismatch between the linguistic representations of length $L$ and the acoustic frames of length $T$, the $\mathbf{H}_{\text{ASP}}$ sequence is utilized along with the Monotonic Alignment Search (MAS) \cite{kim2020glow} (the Length Regulator in Fig.~\ref{fig:architecture}) to quickly learn to align the linguistic representation with the output frames by using the prior loss ($\mathcal{L}_{\text{prior}}$), defined as the negative log-likelihood of the acoustic features given the text prior under the optimal MAS alignment. We route the coarse average spectrogram stream (rather than the source or filter-specific streams) into alignment so that a single, decomposition-agnostic target provides a stable monotonic prior \cite{kim2020glow}. Using these extracted alignments, we obtain the phoneme durations $\mathbf{D} = [d_1, \dots, d_L] \in \mathbb{Z}_{\ge 0}^L$ which we use as ground truth to train the Duration Predictor network~(Fig.~\ref{fig:architecture}),  parameterised by learnable weights $\theta_{\text{dur}}$, which estimates the predicted durations $\hat{\mathbf{D}}$, and is optimised using a logarithmic Mean Squared Error (MSE) loss.

\begin{equation}
    \hat{\mathbf{D}} = \mathcal{F}_{\theta_{\text{dur}}}(\mathbf{H}_{\text{phone}}, \mathbf{e}_s)
    \label{eq:d_hat}
\end{equation}
\vspace{-0.5cm}
\begin{equation}
\mathcal{L}_{\text{dur}} = \frac{1}{L} \sum_{i=1}^{L} \left( \log d_i - \log \hat{d}_i \right)^2
\label{eq:L_dur}
\end{equation}

The ground-truth durations $\mathbf{D}$ are then used to expand $\mathbf{H}_{\text{AEP}}$, $\mathbf{H}_{\text{SEP}}$, and $\mathbf{H}_{\text{ASP}}$ via discrete frame replication to match the target frame length $T$, resulting in the upsampled sequences $\mathbf{H}^*_{\text{AEP}}, \mathbf{H}^*_{\text{SEP}} \in \mathbb{R}^{d_{\text{hidden}} \times T}$ and $\mathbf{H}^*_{\text{ASP}} \in \mathbb{R}^{d_{\text{mel}} \times T}$

\vspace{0.1cm}

\subsubsection{Source and Filter Prediction}
\label{subsubsec:SF_pred}
Following length regulation, the upsampled sequences are routed to their respective physical predictors. The Pitch and Energy Predictor (PEP)~(Fig.~\ref{fig:architecture}), parameterised by learnable weights $\theta_{\text{PEP}}$, estimates the fundamental frequency ($\hat{\mathbf{F}}_0 \in \mathbb{R}^{T}$) and energy ($\hat{\mathbf{E}} \in \mathbb{R}^{T}$) contours:


\begin{equation}
    [\hat{\mathbf{F}}_0, \hat{\mathbf{E}}] = \mathcal{F}_{\theta_{\text{PEP}}}(\mathbf{H}^*_{\text{SEP}}, \mathbf{e}_s)
    \label{eq:f0_e_hat}
\end{equation}

Predicting pitch and energy explicitly exposes the glottal source as a directly controllable signal~\cite{lancucki2021fastpitch, renfastspeech}. The PEP is supervised against the ground-truth pitch ($\mathbf{F}_{0} \in \mathbb{R}^{T}$) and energy ($\mathbf{E} \in \mathbb{R}^{T}$) using a combined MSE and Pearson Correlation Coefficient (PCC) objective:

\begin{equation}
\begin{aligned}
\mathcal{L}_{\text{PE}} &= \frac{1}{T} \left( \|\mathbf{F}_{0} - \hat{\mathbf{F}}_0\|_2^2 + \|\mathbf{E} - \hat{\mathbf{E}}\|_2^2 \right) \\
&\quad + \Big( 2 - \rho(\mathbf{F}_{0}, \hat{\mathbf{F}}_0) - \rho(\mathbf{E}, \hat{\mathbf{E}}) \Big)
\end{aligned}
\label{eq:L_pe}
\end{equation}

where $\rho(\cdot, \cdot)$ denotes the Pearson Correlation Coefficient computed over the sequence length $T$  established in Eq.~\eqref{eq:pcc_def}.

Concurrently, the Articulatory Trajectory Predictor (ATP)~(Fig.~\ref{fig:architecture}), parameterised by learnable weights $\theta_{\text{ATP}}$, processes $\mathbf{H}^*_{\text{AEP}}$ to estimate the continuous kinematic sequence 


\begin{equation}
    \hat{\mathbf{A}} = \mathcal{F}_{\theta_{\text{ATP}}}(\mathbf{H}^*_{\text{AEP}}, \mathbf{e}_s)
    \label{eq:a_hat}
\end{equation}

These predicted trajectories are subsequently supervised against the pseudo-articulatory targets $\tilde{\mathbf{A}}$, which are derived using the AAI model in Section III-A (Eq. \ref{eq:aai_prediction}). Mirroring the PEP objective, we apply the combined MSE loss (Eq. \ref{eq:aai_mse}) and PCC loss (Eq. \ref{eq:pcc}) established in \ref{subsec:AAI}:

\begin{equation}
\mathcal{L}_{\text{ATP}} = \frac{1}{T} \|\tilde{\mathbf{A}} - \hat{\mathbf{A}}\|_F^2 + \Big( 1 - \rho(\tilde{\mathbf{A}}, \hat{\mathbf{A}}) \Big)
\label{eq:L_atp}
\end{equation}

The explicit source-filter disentanglement is achieved during the log-Mel spectrogram generation phase. During training, the Source log-Mel Predictor (SLP)~(Fig.~\ref{fig:architecture}), parameterised by learnable weights $\theta_{\text{SLP}}$, utilises the ground-truth pitch and energy to predict the glottal excitation log-Mel spectrogram, $\hat{\mathbf{M}}_{\text{src}} \in \mathbb{R}^{T \times {d_{\text{mel}}}}$:


\begin{equation}
    \hat{\mathbf{M}}_{\text{src}} = \mathcal{F}_{\theta_{\text{SLP}}}(\mathbf{F}_{0}, \mathbf{E}, \mathbf{H}^*_{\text{ASP}}, \mathbf{e}_s)
    \label{eq:m_src_hat}
\end{equation}

The SLP is conditioned on the ground-truth pitch and energy so that the source spectrogram is explicitly tied to the controllable prosodic contours, while the shared average-spectrogram stream $\mathbf{H}^*_{\text{ASP}}$ supplies the linguistic context needed to render a coherent excitation. Conversely, the Filter log-Mel Predictor (FLP)~(Fig.~\ref{fig:architecture}), parameterised by learnable weights $\theta_{\text{FLP}}$, models the impulse response of the vocal tract by utilising the AAI-generated pseudo-articulatory trajectories to predict the filter spectrogram, $\hat{\mathbf{M}}_{\text{flt}} \in \mathbb{R}^{T \times d_{\text{mel}}}$:


\begin{equation}
    \hat{\mathbf{M}}_{\text{flt}} = \mathcal{F}_{\theta_{\text{FLP}}}(\tilde{\mathbf{A}}, \mathbf{H}^*_{\text{ASP}}, \mathbf{e}_s)
    \label{eq:m_flt_hat}
\end{equation}

The FLP receives the articulatory trajectories, but not the pitch or energy contours. Withholding the source attributes biases the filter to depend only on vocal-tract kinematics, structurally encouraging the source-filter separation. Since the Source log-Mel Predictor (SLP) and Filter log-Mel Predictor (FLP) are explicitly driven by their respective decoupled physical components, pitch and energy for the source, and articulatory trajectories for the filter, they independently estimate the isolated source and filter log-Mel spectrograms. Because multiplication in the linear amplitude domain corresponds to addition in the logarithmic domain, the final synthesised audio log-Mel spectrogram $\hat{\mathbf{M}}_{\text{audio}} \in \mathbb{R}^{T \times d_{\text{mel}}}$ is computed as the direct sum of the source and filter components:

\begin{equation}
\hat{\mathbf{M}}_{\text{audio}} = \hat{\mathbf{M}}_{\text{src}} + \hat{\mathbf{M}}_{\text{flt}}
\label{eq:m_hat_audio}
\end{equation}

This combined prediction is optimised against the ground-truth audio log-Mel spectrogram $\mathbf{M} \in \mathbb{R}^{T \times d_{\text{mel}}}$ using an MSE loss:

\vspace{-0.2cm}
\begin{equation}
\mathcal{L}_{\text{mel}} = \frac{1}{T} \|\mathbf{M} - \hat{\mathbf{M}}_{\text{audio}}\|_F^2
\label{eq:L_mel}
\end{equation}

\subsubsection{Source Refinement via OT-CFM}
\label{subsubsec:src_refin}
Recent work \cite{han2025improving}, has demonstrated that explicitly improving the source excitation component is critical for achieving higher-quality synthesised audio compared to the baseline \cite{bak2021fastpitchformant}. Motivated by this, to further refine the spectral resolution and temporal dynamics of the glottal excitation, we integrate an Optimal Transport Conditional Flow Matching (OT-CFM) module~(Fig.~\ref{fig:architecture}). The MSE-trained source predictor tends to produce over-smoothed spectra. The OT-CFM module restores fine spectral and temporal detail by learning a generative refinement of the source excitation~\cite{lipman2023flow, mehta2024matcha}. To isolate the target source representation for CFM training, we subtract the predicted filter response from the ground-truth audio spectrogram to obtain $\mathbf{M}^{\text{res}}_{\text{src}} = \mathbf{M} - \hat{\mathbf{M}}_{\text{flt}}$.

Let $\mathbf{x}_1 = \mathbf{M}^{\text{res}}_{\text{src}}$ denote the target data sample, and $\mathbf{x}_0 \sim \mathcal{N}(\mathbf{0}, \mathbf{I})$ denote a sample from the prior noise distribution.  We define the aggregated conditioning vector as $\mathbf{c} = 
[\hat{\mathbf{M}}_{\text{src}}; \hat{\mathbf{M}}_{\text{flt}}; \mathbf{e}_s]$. The refinement network is conditioned on the predicted components  ($\hat{\mathbf{M}}_{\text{src}}, \hat{\mathbf{M}}_{\text{flt}}$) so that it refines the source consistently with the already-predicted filter and content rather than in isolation, $\mathbf{e}_s$ preserves speaker identity through the generative step. The OT-CFM module learns to predict the optimal vector field mapping the noise to the target over a continuous time step $\tau \in [0, 1]$. The flow matching loss is formulated as a regression against the target vector field $(\mathbf{x}_1 - \mathbf{x}_0)$:

\begin{equation}
\mathcal{L}_{\text{CFM}} = \mathbb{E}_{\tau \sim \mathcal{U}(0,1), \mathbf{x}_0, \mathbf{x}_1} \left[ \left\| \mathcal{F}_{\theta_{\text{CFM}}}(\mathbf{x}_\tau, \tau, \mathbf{c}) - (\mathbf{x}_1 - \mathbf{x}_0) \right\|_2^2 \right]
\label{eq:L_cfm}
\end{equation}

where $\mathcal{F}_{\theta_{\text{CFM}}}$ is the flow-prediction network with learnable parameters $\theta_{\text{CFM}}$, and $\mathbf{x}_\tau = \tau \mathbf{x}_1 + (1-\tau) \mathbf{x}_0$ represents the interpolated state at time $\tau$.

\subsubsection{Training Objective and Inference}
\label{subsubsec:train_infer}
The entire controllable TTS architecture is optimised end-to-end. The final multi-task training objective is formulated as the direct sum of all individual component penalties:
\begin{equation}
\mathcal{L}_{\text{total}} = \mathcal{L}_{\text{dur}} + \mathcal{L}_{\text{prior}} + \mathcal{L}_{\text{PE}} + \mathcal{L}_{\text{ATP}} + \mathcal{L}_{\text{mel}} + \mathcal{L}_{\text{CFM}}
\end{equation}

where $\mathcal{L}_{\text{dur}}$, $\mathcal{L}_{\text{PE}}$, $\mathcal{L}_{\text{ATP}}$, $\mathcal{L}_{\text{mel}}$, and $\mathcal{L}_{\text{CFM}}$ are defined in Eqs.~\eqref{eq:L_dur}, \eqref{eq:L_pe}, \eqref{eq:L_atp}, \eqref{eq:L_mel}, and \eqref{eq:L_cfm}, respectively, and $\mathcal{L}_{\text{prior}}$ is the MAS alignment prior loss \cite{kim2020glow} introduced earlier in \ref{subsubsec:L_encode_LR}.


At inference, ground-truth acoustic and articulatory targets are unavailable, so the model operates entirely in prediction mode~(the red dashed inference-only paths in Fig.~\ref{fig:architecture}). The intermediate linguistic representations are upsampled using the predicted durations $\hat{\mathbf{D}}$ from Eq.~\eqref{eq:d_hat}. The vocal tract filter response $\hat{\mathbf{M}}_{\text{flt}}$ (Eq.~\eqref{eq:m_flt_hat}) is generated by the FLP conditioned exclusively on the predicted articulatory kinematics $\hat{\mathbf{A}}$ from Eq.~\eqref{eq:a_hat}, in place of the pseudo-targets $\tilde{\mathbf{A}}$ used during training. Correspondingly, the source path is driven by the predicted pitch and energy contours $\hat{\mathbf{F}}_0$ and $\hat{\mathbf{E}}$ from Eq.~\eqref{eq:f0_e_hat}, in place of the ground-truth contours used during training. To reconstruct the glottal excitation, the OT-CFM module samples prior noise $\mathbf{x}_0 \sim \mathcal{N}(\mathbf{0}, \mathbf{I})$ and solves the learned Ordinary Differential Equation (ODE) using an Euler solver, conditioned on the aggregated vector $\mathbf{c}$ defined in Eq.~\eqref{eq:L_cfm}, to generate the refined source spectrogram $\hat{\mathbf{M}}^{\text{res}}_{\text{src}}$. Finally, mirroring the training-phase combination in Eq.~\eqref{eq:m_hat_audio}, the synthesised audio log-Mel spectrogram is obtained via element-wise addition of the two predicted components~(the $\oplus$ node in Fig.~\ref{fig:architecture}):
\begin{equation}
\hat{\mathbf{M}}_{\text{audio}} = \hat{\mathbf{M}}^{\text{res}}_{\text{src}} + \hat{\mathbf{M}}_{\text{flt}}
\end{equation}

This combined spectrogram is subsequently passed to a pre-trained vocoder for final waveform reconstruction.

Overall, the conditioning is asymmetric by design: source-side blocks receive prosodic attributes and filter-side blocks receive articulatory attributes, with only the speaker embedding and a shared linguistic context common to both. This routing encourages the source-filter separation evaluated in Section~\ref{sec:cross-speaker}, and we found it to yield stable training and strong synthesis quality in practice.

\section{Experimental Setup}
\label{sec:exp_setup}
This section details the experimental setup, including the datasets, model architecture, training parameters, and the baseline models used for comparison.

\vspace{-0.3cm}
\subsection{Datasets}
\subsubsection{AAI}
\label{subsubsec:AAI_dataset}

To ensure the model generalises across diverse phonetic profiles, speaking styles, and languages, we utilise a combination of six distinct Electromagnetic Articulography (EMA) datasets. All datasets provide synchronised acoustic waveforms and 12-channel kinematic trajectories, capturing the 2D midsagittal coordinates ($x, y$) of six key articulators: Upper Lip (UL), Lower Lip (LL), Jaw (JAW), Tongue Tip (TT), Tongue Body (TB), and Tongue Dorsum (TD). A comprehensive breakdown of the training corpus, including language demographics, speaker counts, and subset durations, is detailed in Table \ref{tab:AAI_data}.

To assess generalisation, we construct two test sets exclusively from the HPRC dataset: one for intra-speaker performance (seen speakers, unseen utterances) and another for zero-shot inter-speaker evaluation (held-out speakers). Restricting this to HPRC isolates American English, aligning the AAI evaluation with the target dialect of our downstream TTS architecture.

\vspace{-0.1cm}
\begin{table}[htbp]
\centering
\caption{Datasets used for training the AAI model. L1 and L2 denote first (native) and second language, respectively.}
\vspace{-2mm} 
\label{tab:AAI_data}
\resizebox{\columnwidth}{!}{%
\begin{tabular}{@{}llccl@{}}
\toprule
\textbf{Dataset} & \textbf{Language (Accent)} & \textbf{Spks.} & \textbf{Dur.} & \textbf{Description} \\ \midrule
SpireEMA \cite{bandekar2024articulatory} & English (Indian) & 32 & 9h 28m & 460 MOCHA-TIMIT sentences. \\
EMA-MAE \cite{ji2014electromagnetic} & English (American/Mandarin) & 37 & 17h 34m & L1 American \& L2 Mandarin. \\
HPRC \cite{tiede2017quantifying} & English (American) & 6 & 5h 34m & Kinematics of speech rate variation. \\
USC-TIMIT \cite{narayanan2014usc} & English (American) & 2 & 1h 29m & Synchronous core TIMIT sentences. \\
MSPKA \cite{canevari2015new} & Italian (Italian) & 2 & 1h 31m & Hyperarticulated read speech. \\
DKU \cite{cai2018dku} & Cantonese \& Mandarin & 7 & 0h 18m & Native cross-linguistic kinematics. \\ 
\midrule
\textbf{Total} & & \textbf{86} & \textbf{35h 55m} & \\ \bottomrule
\end{tabular}%
}
\vspace{-4mm} 
\end{table}

\subsubsection{TTS}
For the downstream controllable Text-to-Speech (TTS) framework we use the LibriTTS corpus. Training is conducted using the train-clean-460 subset. All audio signals are resampled to a uniform sampling rate of 22.05 kHz.To establish the ground-truth acoustic targets for the model, we compute 80-band log Mel-spectrograms ($d_{\text{mel}} = 80$) as done in \cite{mehta2024matcha} with a hop length of 256 and a window length of 1024. The fundamental frequency ($F_0$) and energy contours are extracted using the PyWORLD vocoder toolkit \footnote{\url{https://github.com/JeremyCCHsu/Python-Wrapper-for-World-Vocoder}}, specifically configured with a matching 256-sample hop length to ensure temporal synchronisation with the Mel-spectrogram acoustic frames. 

For objective evaluation, we report results on the LibriTTS test-clean and test-other subsets. Furthermore, to validate the source-filter-based controllability of the proposed architecture, we utilise the VCTK dataset \cite{yamagishi2019cstr}, leveraging parallel utterances spoken by different speakers to demonstrate cross-speaker control.

\subsection{Model architecture and Training}
\subsubsection{AAI}
The AAI backend~($\mathcal{F}_{\text{AAI}}$, Eq.~\eqref{eq:aai_prediction}) adopts a non-autoregressive encoder-decoder architecture, following the configuration established in \cite{udupa2021estimating}. An initial linear projection layer maps input features into a shared hidden dimension of 256. The core architecture consists of an input encoder and a decoder, each comprising four Feed-Forward Transformer (FFT) blocks that refine the acoustic features and project them into the latent articulatory space. Each FFT block is configured with a single attention head of dimension 32, a convolutional kernel size of 7, a filter size of 256, and a dropout rate of 0.1. A final linear output projection layer maps these hidden states to the 12-channel midsagittal EMA representation. The total parameter count is about 7.3M.

For the acoustic inputs, we investigate contextualised representations extracted from three pre-trained speech models: TERA \cite{liu2021tera} as a self-supervised baseline, MMS-SSL (the base Massively Multilingual Speech model), and MMS-FT (the MMS model fine-tuned for ASR) \cite{pratap2024scaling}. As the different datasets originally sampled articulatory trajectories at varying rates, we downsample all kinematics to a unified 100 Hz. While TERA, which has been used in previous works \cite{udupa2021estimating, bandekar2025enhancing} naturally outputs representations at 100 Hz, the architecture of the MMS models inherently constrains their continuous representations to a 50 Hz sampling rate due to the cumulative stride of their convolutional front-ends. To temporally align these 50 Hz MMS acoustic features with the 100 Hz kinematic data, we apply a discrete frame-duplication strategy, effectively upsampling the acoustic sequence by a factor of two. Finally, each of the 12 kinematic channels is independently normalised using the global mean and standard deviation computed across all utterances within the combined AAI training set.

The model is trained using the Adam optimiser with a fixed learning rate of $1 \times 10^{-4}$. To prevent overfitting during the training process, we implement early stopping monitored against the validation loss. We restrict the maximum audio duration to 30 seconds and train the model with a batch size of 8.

\subsubsection{TTS}
The proposed non-autoregressive encoder-decoder framework utilises a text encoder~($\mathcal{F}_{\theta_{\text{PE}}}$, Eq.~\eqref{eq:h_phone})  and duration predictor~($\mathcal{F}_{\theta_{\text{dur}}}$, Eq.~\eqref{eq:d_hat})  based on Matcha-TTS \cite{mehta2024matcha}. For consistent speaker conditioning, we use the pre-trained speaker embedding module \cite{heo2020clova}, following the configuration of YourTTS \cite{casanova2022yourtts}. The AEP and SEP streams ($\mathcal{F}_{\theta_{\text{AEP}}}$ and $\mathcal{F}_{\theta_{\text{SEP}}}$, Eqs.~\eqref{eq:h_aep} and~\eqref{eq:h_sep})  project linguistic representations to a 256-dimensional hidden space ($d_{\text{hidden}} = 256$) via two sequential 1D-convolutional layers with layer normalisation and ReLU activations. After duplicating these embeddings to the target acoustic length $T$, they are routed to the Pitch and Energy Predictor (PEP)~($\mathcal{F}_{\theta_{\text{PEP}}}$, Eq.~\eqref{eq:f0_e_hat}) and the Articulatory Trajectory Predictor (ATP)~($\mathcal{F}_{\theta_{\text{ATP}}}$, Eq.~\eqref{eq:a_hat}). Both predictors employ 1D-convolutional residual networks with layer normalisation, ReLU, and dropout. To temporally align the modalities, the pseudo-articulatory trajectories are downsampled to match the target temporal length of the mel-spectrogram. The SLP and FLP modules~($\mathcal{F}_{\theta_{\text{SLP}}}$ and $\mathcal{F}_{\theta_{\text{FLP}}}$, Eqs.~\eqref{eq:m_src_hat} and~\eqref{eq:m_flt_hat})  then process these representations using a Transformer architecture adapted from \cite{bak2021fastpitchformant}, consisting of two Feed-Forward Transformer (FFT) blocks. Within these blocks, hidden states are modulated by $\mathbf{e}_s$ via Style-Adaptive Layer Normalisation (SALN) applied after the attention and feed-forward sub-layers.

For acoustic refinement, the OT-CFM flow-prediction network~($\mathcal{F}_{\theta_{\text{CFM}}}$, Eq.~\eqref{eq:L_cfm}) employs a conditional 1D U-Net adapted from \cite{mehta2024matcha}. To estimate the target vector field, it processes a concatenated input comprising the interpolated state $\mathbf{x}_\tau$, predicted source ($\hat{\mathbf{M}}_{\text{src}}$) and filter ($\hat{\mathbf{M}}_{\text{flt}}$) log-Mel spectrograms, and $\mathbf{e}_s$. The continuous time step $\tau$ is encoded via sinusoidal embeddings and dynamically injected into the U-Net. The symmetric down/up-sampling paths use 1D residual networks with Group Normalisation, Mish activations, snake-activated self-attention blocks, and skip connections. A terminal 1D convolution projects these representations to output the predicted vector field, enabling an Euler ODE solver to iteratively transform prior noise into the high-fidelity source spectrogram $\hat{\mathbf{M}}^{\text{res}}_{\text{src}}$. At inference, the OT-CFM source is generated using a Euler ODE solver with 10 steps (NFE\,=\,10), consistent with prior flow-matching TTS systems~\cite{mehta2024matcha}, which report high-quality synthesis at comparable step counts. This refined source is element-wise summed with the predicted filter $\hat{\mathbf{M}}_{\text{flt}}$, and a pre-trained HiFi-GAN vocoder \cite{kong2020hifi} is employed to synthesise the final time-domain audio waveform. The complete TTS architecture comprises approximately 33M parameters and is trained using the Adam optimiser for 1M steps with a batch size of 8, distributed across two 24GB GPUs.

\subsection{Baseline Models}
To rigorously evaluate the proposed AAI-integrated, source-filter architecture, we benchmark its performance against a diverse set of models targeting both articulatory prediction and speech synthesis:

\begin{itemize}
    \item \textbf{Self-Supervised AAI Baselines}: To evaluate the efficacy of our kinematic prediction frontend, we include the TERA SSL \cite{liu2021tera} model as a baseline for Acoustic-to-Articulatory Inversion (AAI), as it has been used before in previous works like \cite{udupa2023improved, bandekar2025enhancing}

    \item \textbf{End-to-End TTS}: We compare against YourTTS \cite{casanova2022yourtts} and XTTS \cite{casanova2024xtts}, which represent highly robust, zero-shot capable, and widely adopted end-to-end architectures. These serve as the standard benchmark for overall speech quality and intelligibility. For both, we use the publicly available open-source code and pre-trained checkpoints.

    \item \textbf{Flow-Matching Models}: Because our acoustic decoder utilises Optimal Transport Conditional Flow Matching (OT-CFM), we include Matcha-TTS \cite{mehta2024matcha} as a direct comparative baseline. To ensure a fair evaluation within our experimental framework, we adapted the standard Matcha-TTS architecture for multi-speaker synthesis by incorporating global speaker embedding conditioning used in YourTTS \cite{casanova2022yourtts}. This adapted baseline was trained for 500,000 updates, allowing us to accurately evaluate the specific impact of our source-filter disentanglement against a rigorously matched, state-of-the-art generative backend.

    \item \textbf{Controllable and Prosodic TTS}: To benchmark our model's explicit prosodic controllability, we compare against StyleTTS2 \cite{li2023styletts} and HierSpeech++ \cite{lee2025hierspeech++}. Both models offer prosodic manipulation and style transfer capabilities, making them the most rigorous competitors for evaluating $F_0$ and energy tracking during our explicit manipulation experiments. For both, we use the publicly available open-source code and pre-trained checkpoints as well.


    \item \textbf{Vanilla Source-Filter Baseline:} To isolate the contribution of articulatory conditioning, we include an ablation of the proposed architecture with the entire articulatory branch removed (the AEP, ATP, and Filter log-Mel Predictor's articulatory input). The resulting baseline retains the source-side components (SEP, PEP, Source log-Mel Predictor, OT-CFM) and predicts the filter log-Mel from $\mathbf{H}^*_{\text{ASP}}$ and the speaker embedding alone, mirroring the source-filter decomposition of [8] but adapted to use an OT-CFM source decoder for fair comparison. This isolates the contribution of physical articulatory grounding from the rest of the architectural choices.

     \item \textbf{No-OT-CFM Ablation:} To quantify the contribution of the flow-matching refinement, we evaluate a variant of the proposed model in which the OT-CFM module is removed and the source spectrogram is taken directly from the MSE-trained source predictor. All other components are unchanged. We report objective metrics for this ablation; subjective MOS was not collected for this condition.

\end{itemize}


\begin{table}[t]
\centering
\renewcommand{\arraystretch}{0.95} 
\caption{CC for articulatory trajectory prediction on the HPRC test set. The proposed TTS model is zero-shot for both speaker subsets. Best values bolded.}
\label{tab:AAI_perform}
\begin{tabular}{clcc}
\toprule
\textbf{Type} & \textbf{Model} & \textbf{Seen Speakers} & \textbf{Unseen Speakers} \\ 
\midrule
\multirow{3}{*}{AAI} & TERA & 0.9153 & 0.7226 \\ 
 & MMS-SSL & 0.9450 & \textbf{0.7749} \\ 
 & MMS-FT & \textbf{0.9542} & 0.7711 \\ 
\midrule
TTS & Proposed (MMS-FT) & 0.6673 & 0.6922 \\ 
\bottomrule
\end{tabular}
\vspace{-4mm}
\end{table}


\begin{table*}[t]
\centering
\renewcommand{\arraystretch}{0.95}
\setlength{\tabcolsep}{4pt}
\caption{Prosodic tracking and spectral fidelity on the LibriTTS test sets. Params is the inference-time count (excluding vocoder and speaker-embedding modules). Ctrl.\ and Int.\ are additive scores over exposed controllable and interpretable representations; see Section~\ref{subsec:score_metrics} for definitions. Best values per column in bold.}
\vspace{-0.2cm}
\label{tab:TTS_prosody}
\begin{tabular}{l c cc cc cc cc}
\toprule
\multirow{2}{*}{\textbf{Model}} & \multirow{2}{*}{\textbf{Params (M) $\downarrow$}} & \multirow{2}{*}{\textbf{Ctrl. $\uparrow$}} & \multirow{2}{*}{\textbf{Int. $\uparrow$}} & \multicolumn{2}{c}{\textbf{Log F0 MSE $\downarrow$}} & \multicolumn{2}{c}{\textbf{F0 CC $\uparrow$}} & \multicolumn{2}{c}{\textbf{MCD $\downarrow$}} \\
\cmidrule(lr){5-6} \cmidrule(lr){7-8} \cmidrule(lr){9-10}
 & & \textbf{(/3)} & \textbf{(/5)} & \textbf{clean} & \textbf{other} & \textbf{clean} & \textbf{other} & \textbf{clean} & \textbf{other} \\
\midrule
Matcha-TTS & \textbf{20.9} & 0 & 0 & 1.15 & 1.62 & 0.8797 & 0.8013 & 5.67 & 5.99 \\
StyleTTS2 & 129.5 & 2 & 2 & 1.05 & 1.36 & 0.8954 & 0.8221 & \textbf{4.52} & \textbf{4.80} \\
YourTTS & 22.0 & 0 & 0 & 1.36 & 1.90 & 0.8718 & 0.7881 & 5.69 & 5.77 \\
XTTS & 466.9 & 0 & 0 & 1.38 & 2.01 & 0.8730 & 0.7908 & 5.67 & 5.77 \\
HierSpeech++ & 205.6 & 1 & 1 & \textbf{0.96} & \textbf{1.22} & \textbf{0.9071} & \textbf{0.8429} & 4.91 & 5.35 \\
\midrule
Source-Filter & 30.0 & 2 & 4 & 1.17 & 1.61 & 0.8815 & 0.7993 & 5.73 & 6.07 \\
Proposed (MMS-FT) & 33.0 & \textbf{3} & \textbf{5} & 1.19 & 1.63 & 0.8841 & 0.8004 & 5.89 & 6.36 \\
\quad \textit{w/o OT-CFM} & 33.0 & \textbf{3} & \textbf{5} & 1.76 & 2.25 & 0.8351 & 0.7392 & 6.72 & 7.49 \\
\bottomrule
\end{tabular}
\vspace{-4mm}
\end{table*}

\begin{table*}[t]
\centering
\setlength{\tabcolsep}{3.5pt}
\renewcommand{\arraystretch}{0.95}
\caption{Objective and subjective evaluation of speech intelligibility, speaker similarity, and naturalness on the LibriTTS test sets. Params is the inference-time count (excluding vocoder and speaker-embedding modules); Ctrl.\ and Int.\ are additive controllability/interpretability scores (see Section~\ref{subsec:score_metrics}). Best performing values per column are in bold.}
\vspace{-0.2cm}
\label{tab:TTS_obj_sub}
\resizebox{1.0\textwidth}{!}{
\begin{tabular}{l c cc cc cc cc cc cc}
\toprule
\multirow{2}{*}{\textbf{Model}} & \multirow{2}{*}{\textbf{Params (M) $\downarrow$}} & \multirow{2}{*}{\textbf{Ctrl. $\uparrow$}} & \multirow{2}{*}{\textbf{Int. $\uparrow$}} & \multicolumn{2}{c}{\textbf{WER ($\downarrow$)}} & \multicolumn{2}{c}{\textbf{CER ($\downarrow$)}} & \multicolumn{2}{c}{\textbf{Speaker Similarity ($\uparrow$)}} & \multicolumn{2}{c}{\textbf{MOS ($\uparrow$)}} & \multicolumn{2}{c}{\textbf{UTMOS ($\uparrow$)}} \\ 
\cmidrule(lr){5-6} \cmidrule(lr){7-8} \cmidrule(lr){9-10} \cmidrule(lr){11-12} \cmidrule(lr){13-14}
 & & \textbf{(/3)} & \textbf{(/5)} & \textbf{clean} & \textbf{other} & \textbf{clean} & \textbf{other} & \textbf{clean} & \textbf{other} & \textbf{clean} & \textbf{other} & \textbf{clean} & \textbf{other} \\ 
\midrule
Matcha TTS & \textbf{20.9} & 0 & 0 & 4.28 & \textbf{5.39} & \textbf{1.63} & \textbf{2.25} & 0.9418 & 0.9089 & 3.93 $\pm$ 0.14 & 3.85 $\pm$ 0.14 & 3.66 $\pm$ 0.012 & 3.44 $\pm$ 0.012 \\ 
StyleTTS2 & 129.5 & 2 & 2 & \textbf{4.17} & 5.40 & 2.43 & 3.98 & 0.9408 & 0.9081 & \textbf{4.17 $\pm$ 0.13} & 4.16 $\pm$ 0.12 & 4.09 $\pm$ 0.010 & 3.84 $\pm$ 0.013 \\ 
YourTTS & 22.0 & 0 & 0 & 8.51 & 10.20 & 3.74 & 4.74 & 0.9383 & 0.8861 & 3.63 $\pm$ 0.15 & 3.59 $\pm$ 0.16 & 3.59 $\pm$ 0.012 & 3.48 $\pm$ 0.013 \\ 
XTTS & 466.9 & 0 & 0 & 5.52 & 8.35 & 2.93 & 4.79 & 0.9257 & 0.9049 & 3.93 $\pm$ 0.15 & 3.45 $\pm$ 0.17 & 3.65 $\pm$ 0.016 & 3.27 $\pm$ 0.018 \\ 
HierSpeech++ & 205.6 & 1 & 1 & 4.65 & 6.13 & 1.92 & 2.72 & \textbf{0.9489} & \textbf{0.9103} & 4.16 $\pm$ 0.14 & \textbf{4.31 $\pm$ 0.13} & \textbf{4.23 $\pm$ 0.009} & \textbf{4.09 $\pm$ 0.011} \\ 
\midrule
Source-Filter & 30.0 & 2 & 4 & 5.96 & 7.96 & 2.66 & 3.76 & 0.9333 & 0.9006 & 3.74 $\pm$ 0.15 & 3.77 $\pm$ 0.16 & 3.75 $\pm$ 0.011 & 3.56 $\pm$ 0.012 \\ 
Proposed (MMS-FT) & 33.0 & \textbf{3} & \textbf{5} & 5.53 & 7.33 & 2.29 & 3.35 & 0.9366 & 0.8989 & 3.76 $\pm$ 0.15 & 3.65 $\pm$ 0.16 & 3.71 $\pm$ 0.011 & 3.53 $\pm$ 0.012 \\ 
\quad \textit{w/o OT-CFM} & 33.0 & \textbf{3} & \textbf{5} & 4.77 & 6.40 & 1.91 & 2.87 & 0.9309 & 0.8875 & --$^{\dagger}$ & --$^{\dagger}$ & 2.26 $\pm$ 0.014 & 1.99 $\pm$ 0.014 \\
\midrule
\multicolumn{14}{@{}l}{\footnotesize $^{\dagger}$ Subjective MOS was not collected for the no-OT-CFM ablation; only objective metrics are reported.} \\
\bottomrule
\end{tabular}
}
\vspace{-4mm}
\end{table*}

\section{Evaluation Metrics}
\label{sec:metrics}
To comprehensively evaluate the proposed source-filter based architecture, we employ a diverse set of objective and subjective metrics targeting articulatory precision, acoustic quality, physical interpretability and controllability.

\subsection{Articulatory Trajectory Prediction Performance}
To validate the physical accuracy of the kinematic representations throughout the pipeline, we conduct evaluations using ground-truth EMA data from the HPRC test set with both seen and unseen speakers. We quantify precision using the Pearson Correlation Coefficient (CC) as done in previous AAI works. This evaluation is performed at two critical stages of the framework:

\begin{itemize}
    \item \textbf{Acoustic-to-Articulatory Inversion (AAI)}: We benchmark the inversion accuracy across three contextualised acoustic representations (TERA, MMS-SSL, and MMS-FT) to determine the optimal input feature space for kinematic prediction.

    \item \textbf{Text-to-Articulatory Prediction}: To ensure the downstream TTS model learns valid human articulatory behaviour rather than just overfitting to pseudo-targets, we explicitly evaluate the zero-shot physical accuracy of the text-predicted trajectories against the HPRC test data.
\end{itemize}

\subsection{Speech Synthesis and Model Evaluation}
\label{subsec:score_metrics}
The downstream TTS performance is evaluated across several dimensions:

\begin{itemize}
    \item \textbf{Intelligibility}: WER and CER via a pre-trained whisper-medium ASR\footnote{\url{https://huggingface.co/openai/whisper-medium}}.

    \item \textbf{Spectral and Prosodic Fidelity}: MCD for spectra; log-MSE and CC for F0 and energy contours.

    \item \textbf{Speaker Similarity}: Cosine similarity of WavLM speaker embeddings\footnote{\url{https://huggingface.co/microsoft/wavlm-base-sv}}.

     \item \textbf{Naturalness}: Subjective naturalness is evaluated via MOS listening tests: 20 listeners\footnote{Informed consent was obtained from all participants involved in the subjective listening evaluations.} rated 20 utterances per system (10 test-clean, 10 test-other) on a 5-point scale from 1 (completely unnatural) to 5 (very natural). To complement these subjective human evaluations with a scalable, automated measure, we additionally employ UTMOS \cite{saeki2022utmos}, a neural network-based naturalness predictor, computed over the full LibriTTS test-clean and test-other sets. 

     \item \textbf{Controllability and Interpretability Scores}: We quantify each model as additive scores. The controllability score (max~3) adds one point per explicitly manipulable attribute (pitch, energy, articulatory trajectories); the interpretability score (max~5) adds one point per exposed intermediate representation (pitch, energy, articulatory contours, and source and filter spectrograms).
     
\end{itemize}

\begin{table*}[t]
\centering
\renewcommand{\arraystretch}{0.95} 
\caption{Performance of explicit source attribute manipulation across various scaling factors. Params (M) denotes the inference-time parameter count (excluding vocoder and speaker-embedding modules). Metrics report the deviation and correlation against ground-truth manipulated contours. Best performing values per scale are in bold.}
\label{tab:prosodic_scaling}
\resizebox{\textwidth}{!}{
\begin{tabular}{lc cc cc cc cc cc c}
\toprule
\multirow{2}{*}{\textbf{Model}} & \multirow{2}{*}{\textbf{Scale}} & \multicolumn{2}{c}{\textbf{F0}} & \multicolumn{2}{c}{\textbf{Energy}} & \multicolumn{2}{c}{\textbf{F1}} & \multicolumn{2}{c}{\textbf{F2}} & \multicolumn{2}{c}{\textbf{F3}} & \multirow{2}{*}{\textbf{MCD}} \\
\cmidrule(lr){3-4} \cmidrule(lr){5-6} \cmidrule(lr){7-8} \cmidrule(lr){9-10} \cmidrule(lr){11-12}
 & & \textbf{Log MSE} & \textbf{CC} & \textbf{Log MSE} & \textbf{CC} & \textbf{Log MSE} & \textbf{CC} & \textbf{Log MSE} & \textbf{CC} & \textbf{Log MSE} & \textbf{CC} & \\
\midrule
\multirow{6}{*}{\shortstack[l]{\textbf{Proposed}\\\footnotesize (\textbf{33.0M})}} 
 & 0.7 & 2.2393 & 0.7498 & \textbf{0.0287} & \textbf{0.9859} & 0.0848 & \textbf{0.8839} & 0.0173 & 0.8930 & 0.0068 & 0.8605 & 5.0352 \\
 & 0.8 & 2.1528 & 0.7424 & \textbf{0.0256} & \textbf{0.9852} & 0.0778 & 0.8807 & 0.0179 & 0.8911 & 0.0074 & 0.8555 & 4.8292 \\
 & 0.9 & 2.2279 & 0.7673 & \textbf{0.0245} & \textbf{0.9890} & 0.0781 & 0.8849 & 0.0175 & 0.8971 & 0.0130 & 0.8583 & 4.6688 \\
 & 1.1 & 2.1807 & 0.7484 & \textbf{0.0255} & \textbf{0.9875} & 0.0768 & 0.8827 & 0.0171 & 0.8960 & 0.0078 & 0.8557 & 4.6850 \\
 & 1.2 & 2.2844 & 0.7500 & \textbf{0.0258} & \textbf{0.9869} & 0.0747 & \textbf{0.8890} & 0.0165 & 0.8995 & 0.0073 & 0.8669 & 4.9371 \\
 & 1.3 & 2.1259 & 0.7425 & \textbf{0.0284} & \textbf{0.9849} & 0.0742 & 0.8772 & 0.0182 & 0.8928 & 0.0076 & 0.8559 & 5.0993 \\
\midrule
\multirow{6}{*}{\shortstack[l]{\textbf{StyleTTS2}\\\footnotesize (129.5M)}} 
 & 0.7 & \textbf{1.9301} & \textbf{0.8175} & 0.2289 & 0.8212 & \textbf{0.0795} & 0.8760 & \textbf{0.0165} & \textbf{0.9052} & 0.0081 & \textbf{0.8796} & \textbf{3.5431} \\
 & 0.8 & \textbf{1.8092} & \textbf{0.8143} & 0.1016 & 0.9111 & \textbf{0.0672} & \textbf{0.8901} & \textbf{0.0146} & \textbf{0.9191} & 0.0065 & \textbf{0.8960} & \textbf{2.8277} \\
 & 0.9 & \textbf{1.8747} & \textbf{0.8374} & 0.0735 & 0.9551 & \textbf{0.0637} & \textbf{0.8944} & \textbf{0.0140} & \textbf{0.9180} & 0.0069 & \textbf{0.8984} & \textbf{2.7099} \\
 & 1.1 & \textbf{1.8007} & \textbf{0.8443} & 0.0643 & 0.9603 & \textbf{0.0580} & \textbf{0.8883} & \textbf{0.0135} & \textbf{0.9220} & 0.0066 & \textbf{0.8998} & \textbf{2.4724} \\
 & 1.2 & \textbf{1.8420} & \textbf{0.8289} & 0.0692 & 0.9529 & \textbf{0.0635} & 0.8785 & \textbf{0.0146} & \textbf{0.9104} & \textbf{0.0058} & \textbf{0.8853} & \textbf{3.2053} \\
 & 1.3 & \textbf{1.8571} & \textbf{0.8173} & 0.0789 & 0.9454 & \textbf{0.0666} & 0.8567 & \textbf{0.0150} & \textbf{0.9096} & \textbf{0.0062} & \textbf{0.8818} & \textbf{3.2337} \\
\midrule
\multirow{6}{*}{\shortstack[l]{\textbf{HierSpeech++}\\\footnotesize (205.6M)}} 
 & 0.7 & 3.2795 & 0.7584 & --$^{\dagger}$ & --$^{\dagger}$ & 0.0930 & 0.8753 & 0.0182 & 0.8759 & \textbf{0.0062} & 0.8543 & 3.6290 \\
 & 0.8 & 2.9639 & 0.7630 & -- & -- & 0.0853 & 0.8867 & 0.0174 & 0.8782 & \textbf{0.0064} & 0.8587 & 3.1209 \\
 & 0.9 & 2.9638 & 0.7614 & -- & -- & 0.0939 & 0.8710 & 0.0189 & 0.8659 & \textbf{0.0065} & 0.8506 & 3.1049 \\
 & 1.1 & 2.9711 & 0.7643 & -- & -- & 0.0825 & 0.8848 & 0.0175 & 0.8783 & \textbf{0.0065} & 0.8619 & 2.8088 \\
 & 1.2 & 2.9124 & 0.7639 & -- & -- & 0.0884 & 0.8812 & 0.0181 & 0.8747 & 0.0067 & 0.8553 & 3.3561 \\
 & 1.3 & 3.0763 & 0.7440 & -- & -- & 0.0874 & \textbf{0.8796} & 0.0181 & 0.8745 & 0.0064 & 0.8578 & 3.2784 \\
 \midrule
 \multicolumn{13}{l}{\footnotesize $^{\dagger}$ Energy manipulation is not applicable to HierSpeech++, which supports only explicit F0 control.} \\
\bottomrule
\end{tabular}
}
\vspace{-4mm}
\end{table*}

\begin{table*}[t]
\centering
\renewcommand{\arraystretch}{0.95} 
\caption{Cross-speaker disentanglement evaluation. The table reports acoustic tracking metrics when generating speech using varying combinations of source excitations and vocal tract kinematic filters from Speaker A and Speaker B.}
\label{tab:disentanglement}
\setlength{\tabcolsep}{3.5pt}
\begin{tabular}{ll cc cc cc cc cc c}
\toprule
\multirow{2}{*}{\textbf{Scenario}} & \multirow{2}{*}{\begin{tabular}[c]{@{}l@{}}\textbf{Evaluation}\\ \textbf{Target}\end{tabular}} & \multicolumn{2}{c}{\textbf{F0}} & \multicolumn{2}{c}{\textbf{Energy}} & \multicolumn{2}{c}{\textbf{F1}} & \multicolumn{2}{c}{\textbf{F2}} & \multicolumn{2}{c}{\textbf{F3}} & \multirow{2}{*}{\textbf{MCD}} \\
\cmidrule(lr){3-4} \cmidrule(lr){5-6} \cmidrule(lr){7-8} \cmidrule(lr){9-10} \cmidrule(lr){11-12}
 & & \textbf{Log MSE} & \textbf{CC} & \textbf{Log MSE} & \textbf{CC} & \textbf{Log MSE} & \textbf{CC} & \textbf{Log MSE} & \textbf{CC} & \textbf{Log MSE} & \textbf{CC} & \\
\midrule
\multirow{2}{*}{\begin{tabular}[c]{@{}l@{}}\textbf{A-B}\\ \textbf{(Cross-Speaker)}\end{tabular}} 
 & Speaker A & 1.5988 & 0.7587 & 0.4259 & 0.8414 & 0.1602 & 0.7777 & 0.0327 & 0.7938 & 0.0146 & 0.7650 & 5.8321 \\
 & Speaker B & 1.9843 & 0.6850 & 0.4597 & 0.8349 & 0.1578 & 0.7791 & 0.0303 & 0.8004 & 0.0134 & 0.7769 & 5.4694 \\
\midrule
\multirow{2}{*}{\begin{tabular}[c]{@{}l@{}}\textbf{A-A}\\ \textbf{(Speaker A Only)}\end{tabular}} 
 & Speaker A & 1.5757 & 0.7663 & 0.4189 & 0.8457 & 0.1551 & 0.7883 & 0.0312 & 0.8052 & 0.0133 & 0.7837 & 5.4056 \\
 & Speaker B & 1.9569 & 0.6872 & 0.4636 & 0.8319 & 0.1642 & 0.7756 & 0.0333 & 0.7890 & 0.0144 & 0.7649 & 5.7776 \\
\midrule
\multirow{2}{*}{\begin{tabular}[c]{@{}l@{}}\textbf{B-B}\\ \textbf{(Speaker B Only)}\end{tabular}} 
 & Speaker A & 1.8584 & 0.6922 & 0.4359 & 0.8349 & 0.1646 & 0.7705 & 0.0340 & 0.7885 & 0.0154 & 0.7610 & 6.1317 \\
 & Speaker B & 1.6522 & 0.7602 & 0.4371 & 0.8392 & 0.1556 & 0.7783 & 0.0305 & 0.8024 & 0.0135 & 0.7811 & 5.2231 \\
\bottomrule
\end{tabular}
\vspace{-4mm}
\end{table*}

\subsection{Explicit Control and Disentanglement Analysis}

To quantify explicit controllability and structural disentanglement, we perform prosodic manipulation on the VCTK dataset. We artificially scale the predicted pitch and energy contours by multipliers ranging from 0.7 to 1.3 during synthesis. We evaluate the generated $F_0$, energy, and formants ($F_1$, $F_2$, $F_3$) against equivalently scaled ground-truths using log-MSE, Pearson Correlation (CC), and Mel Cepstral Distortion (MCD)\footnote{In MCD calculations, the zeroth coefficient ($c_0$) is excluded to decouple overall frame energy from the spectral envelope evaluation.}. A perfectly disentangled model tracks the scaled source parameters while leaving filter characteristics ($F_{1-3}$) unperturbed. We benchmark this explicit control against StyleTTS2 and HierSpeech++.

To further validate physical interpretability, we conduct a cross-speaker source-filter disentanglement analysis using identical utterances from two speakers (A and B). We explicitly divide the speaker embedding routing: Speaker A's embedding is injected into all source and temporal modules, while Speaker B's drives all filter blocks (the A-B setup). Effective disentanglement is demonstrated if the A-B synthesis strongly tracks Speaker A's prosody ($F_0$, energy) and Speaker B's vocal-tract resonances ($F_{1-3}$). To rigorously verify parameter isolation, we benchmark these cross-speaker correlations against self-reconstructed (A-A and B-B) baselines.

Furthermore, we conduct a qualitative graphical analysis of the articulatory-acoustic relationship to prove physical grounding:

\begin{itemize}
    \item \textbf{Kinematic Disentanglement in Minimal Pairs:} We analyse predicted articulatory trajectories for the minimal pair ``pat'' and ``cat'' to verify the model's ability to correctly isolate primary articulators (lips vs. tongue dorsum) while preserving shared coarticulatory trends.
    \item \textbf{Acoustic-Kinematic Realisation:} We evaluate the natural mapping of acoustic signals to complex physical postures by tracing the anticipatory retroflex gesture and the simultaneous $F3$ lowering characteristic of the American rhotic /r/ in the word ``cart''.
    \item \textbf{Controllability via Trajectory Modification:} We demonstrate fine-grained dialectal control by manually flattening the tongue elevation trajectories in the word ``bird''. This direct physical intervention suppresses acoustic rhoticity.
\end{itemize}

\section{Results and Discussion}
\label{sec:results}
In this section, we present the results of the evaluations detailed in Section \ref{sec:metrics}\footnote{Audio samples, source/filter decompositions, and the articulatory-edit examples are available at \url{https://coding-phoenix-12.github.io/ArticulatorySFTTS/}}.

\vspace{-0.3cm}
\subsection{AAI Backend Performance}
\label{results}

Table~\ref{tab:AAI_perform} reports the Pearson Correlation Coefficient (CC) on the HPRC test set for both seen and unseen speakers. We evaluate three feature-extraction configurations of our AAI backend (TERA, MMS-SSL, and ASR-finetuned MMS, MMS-FT). MMS-FT achieves the highest correlation for seen speakers (0.9542), while MMS-SSL marginally leads on unseen speakers (0.7749). We adopt MMS-FT as the feature extractor for downstream pseudo-labelling, as preliminary experiments showed it yielded better downstream TTS performance than MMS-SSL under matched training conditions.

The bottom row of Table~\ref{tab:AAI_perform} reports the Text-to-Articulatory Trajectory Evaluation: the zero-shot articulatory accuracy of the proposed TTS model's ATP block on the HPRC test set. The proposed TTS model is trained on MMS-FT pseudo-labels using only LibriTTS, without any exposure to HPRC. Since every HPRC speaker is effectively unseen from the TTS model's perspective, the relevant comparison is against the unseen-speaker performance of the dedicated AAI baselines. Under this comparison, the proposed TTS achieves a CC of 0.6922 on unseen speakers, close to the dedicated MMS-FT baseline's unseen-speaker performance (0.7711) despite the added challenges of pseudo-label supervision, phoneme rather than acoustic input, and cross-corpus generalisation, indicating that the model learns valid articulatory dynamics from text under indirect supervision.

\subsection{TTS Quality and Model Capabilities}

Table~\ref{tab:TTS_obj_sub} details downstream performance on the LibriTTS subsets. The proposed kinematic-conditioned model (MMS-FT) achieves intelligibility comparable to or slightly better than the Vanilla Source-Filter baseline across both sets (WER 5.53 vs.\ 5.96 on test-clean), while maintaining competitive naturalness (MOS 3.76, UTMOS 3.71), suggesting that articulatory conditioning provides useful structure beyond what is captured by speaker embedding and pitch/energy contours alone. Although removing the OT-CFM refinement yields marginally better intelligibility (WER 4.77 vs.\ 5.53, CER 1.91 vs.\ 2.29 on test-clean), it severely degrades naturalness and spectral fidelity (UTMOS 3.71 to 2.26, MCD 5.89
to 6.72), confirming that the MSE-trained source alone is over-smoothed and that the flow-matching step is essential for restoring perceptual quality.

These results must be contextualised by model capacity. While massive, unconstrained models like StyleTTS2 and HierSpeech++ achieve higher raw MOS and speaker similarity, they utilise significantly larger architectures. Despite its smaller footprint, our model matches the far larger XTTS in intelligibility (WER 5.53 vs.\ 5.52), outperforms the comparably sized YourTTS across all perceptual metrics, and is highly competitive with Matcha-TTS. This confirms that the explicit source-filter decomposition delivers competitive quality while providing interpretability and disentanglement that black-box baselines lack as summarised by the controllability and interpretability scores in Tables~\ref{tab:TTS_obj_sub} and~\ref{tab:TTS_prosody}. The proposed model is the only system to attain the maximum on both, exposing all three controllable attributes and both spectral decompositions, whereas the black-box baselines score zero.

\subsection{Prosodic and Spectral Fidelity}

Table \ref{tab:TTS_prosody} details prosodic tracking and spectral fidelity. The proposed architecture achieves competitive F0 reconstruction (CC 0.8841 on test-clean), exceeding both XTTS (0.8730) and YourTTS (0.8718), with the trend consistent on test-other. Its MCD (5.89) is higher than the much larger StyleTTS2 (4.52) and HierSpeech++ (4.91), an expected consequence of constraining the output to the strict source-filter decomposition.

\begin{figure}[t]
    \centering
    \includegraphics[width=0.85\linewidth]{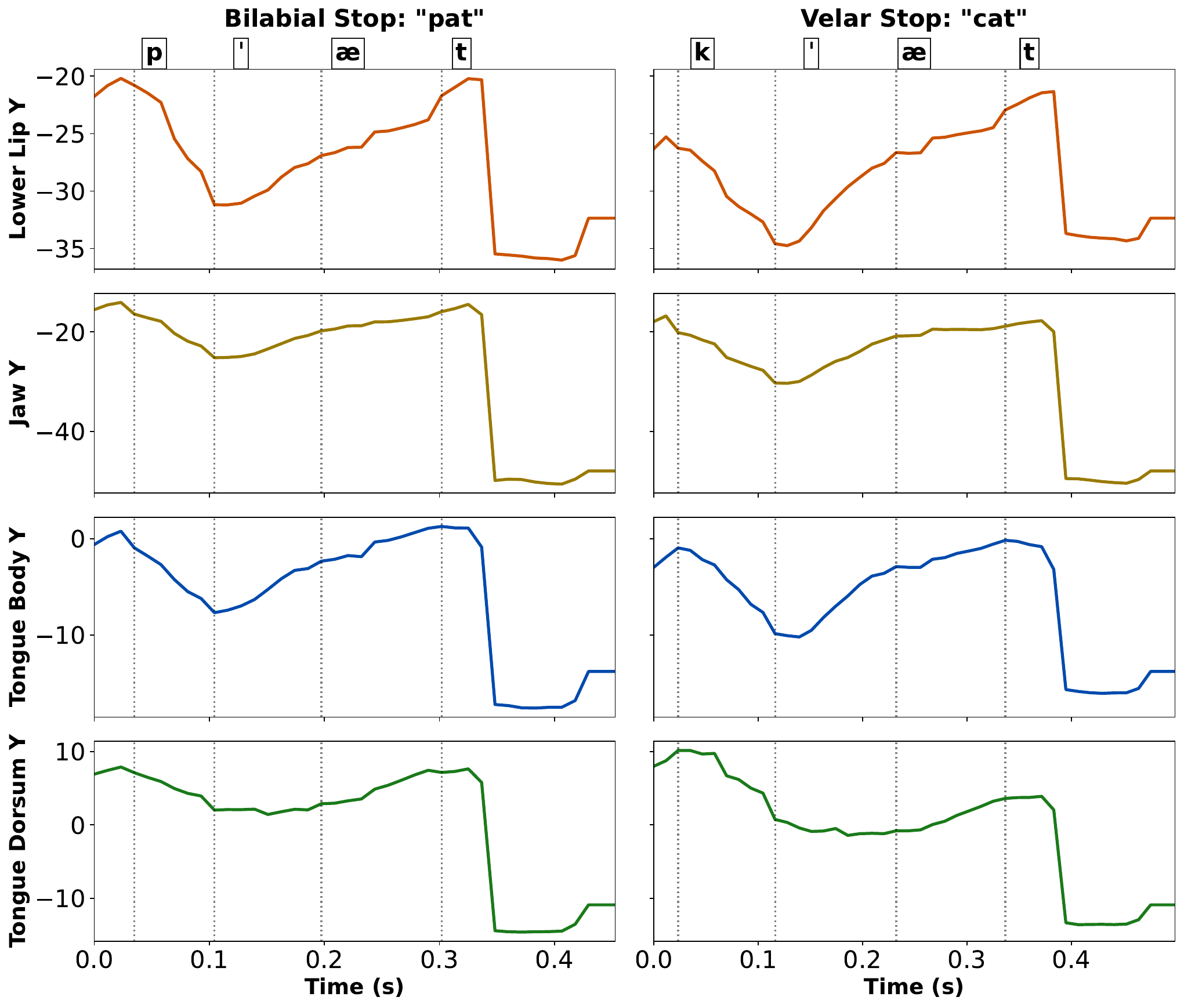}
    \vspace{-3mm}
    \caption{Kinematic trajectory comparison for the minimal pair ``pat'' (bilabial stop) and ``cat'' (velar stop), showing generated vertical displacement of the lower lip, jaw, tongue body, and tongue dorsum. Dashed lines mark phoneme boundaries for the stop closure, the vowel /\ae/, and the final /t/.}
    \label{fig:minimal_pair_kinematics}
      \vspace{-4mm}
\end{figure}

\begin{figure}[htbp]
    \centering
    \includegraphics[width=0.85\linewidth]{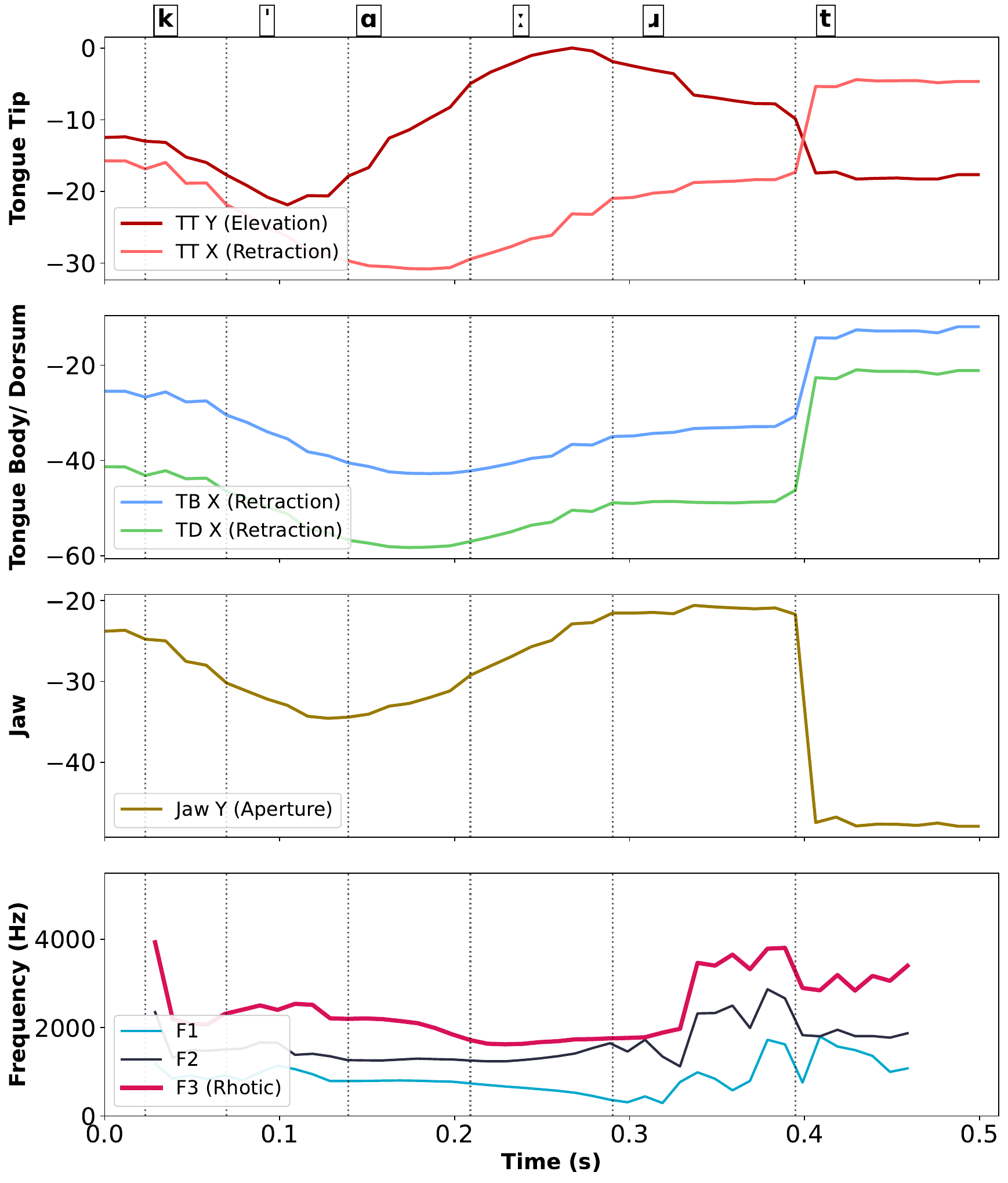}
    \vspace{-3mm}
    \caption{Acoustic and kinematic trajectories for the word ``cart'' (/k'a:rt/). The acoustic panel (bottom) shows the characteristic F3 lowering at 0.20--0.30s; the kinematic panels show the corresponding retroflex gesture via simultaneous tongue-tip elevation ($TT_{y}$) and retraction ($TT_{x}$) decoupled from the lowered jaw ($Jaw_{y}$)}
    \label{fig:cart_rhotic}
  \vspace{-6mm}
\end{figure}

\begin{figure*}[htbp]
    \centering
    \includegraphics[width=0.70\textwidth]{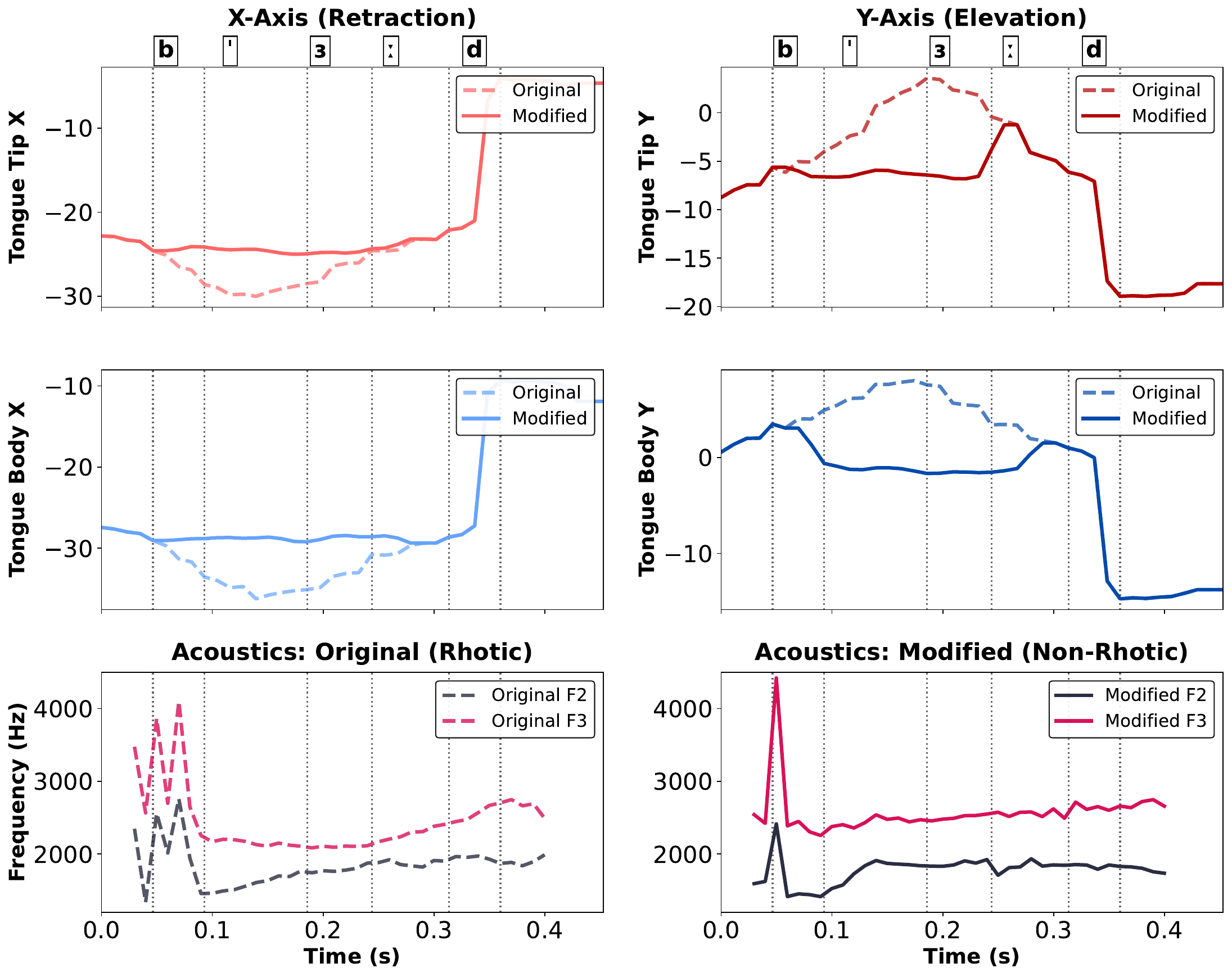}
    \vspace{-5mm}
    \caption{Kinematic and acoustic effects of modifying the rhotic vowel in ``bird''. Top two rows: Tongue Tip ($TT_y$) and Tongue Body ($TB_y$) elevation. Flattening the original bunched/retroflex gesture (dashed$\rightarrow$solid) shifts the acoustic synthesis (bottom) from a suppressed rhotic $F3$ (left) to a non-rhotic $F3$ plateau (right), preventing $F2$--$F3$ convergence.}
    \label{fig:bird_modification}
      \vspace{-4mm}
\end{figure*}

\subsection{Explicit Source Attributes Controllability}

To evaluate the robustness and disentanglement of the proposed architecture, we conducted explicit manipulation of the source attributes. Specifically, the fundamental frequency ($F_0$) and energy contours were uniformly scaled by factors ranging from 0.7 to 1.3 during inference. It is important to note that HierSpeech++ natively supports only explicit $F_0$ manipulation; therefore, energy manipulation results are not reported for HierSpeech++. The generated audio was then analysed to determine how accurately the synthesised pitch and energy tracked the target scaled contours, and to measure the unintended degradation on the vocal tract filter response, quantified by formant tracking ($F_1, F_2, F_3$) and Mel-Cepstral Distortion (MCD). The results are detailed in Table \ref{tab:prosodic_scaling}.

The proposed model demonstrates better stability and precision in explicit energy manipulation. Across all scaling factors it maintains a near-perfect Energy CC (0.9849–0.9890) and the lowest Log MSE. By contrast, StyleTTS2 performs well near the neutral scale but degrades as scaling moves further from the training distribution, showing that the proposed architecture provides granular prosodic control with greater stability under extreme scaling.

For pitch, the larger StyleTTS2 baseline achieves the best raw F0 correlation, while the proposed model trails StyleTTS2 but exceeds HierSpeech++ across all scales, indicating that the gap reflects model capacity rather than a disentanglement failure. However, the formant tracking remains on par with the baselines ($F_1$ CC $\approx$ 0.88), indicating that the manipulated source passes through the kinematic-conditioned filter without disturbing the predicted vocal-tract resonances.

\subsection{Cross-Speaker Disentanglement}
\label{sec:cross-speaker}

Table \ref{tab:disentanglement} shows that the proposed architecture largely isolates the excitation source from the kinematic vocal tract filter.

The clearest evidence of disentanglement is in the pitch contour. In the cross-speaker (A-B) setup, the generated audio correlates strongly with Speaker A's ground-truth F0 (CC: 0.7587), closely approaching the A-A self-reconstruction ceiling (0.7663). Its correlation with Speaker B's F0 (0.6850) is comparable to the B-B → Speaker A baseline (0.6922), confirming that no spurious pitch tracking occurs and that the F0 contour remains strictly tied to the source embedding. This tracking behaviour inverts cleanly for spectral characteristics, with A-B output formants (F1, F2, F3) aligning more closely with Speaker B than Speaker A. Mel-Cepstral Distortion (MCD), which evaluates the spectral envelope independently of excitation, further confirms this structural transfer: the A-B synthesis achieves a lower (better) MCD against Speaker B (5.4694) than against Speaker A (5.8321), a gap comparable to the A-A self-reconstruction's spectral preference, indicating partial but consistent transfer of Speaker B's spectral characteristics through the filter routing. Energy follows the same direction (A-B → Speaker A: 0.8414, close to the A-A ceiling of 0.8457), though the speakers' energy contours are similar enough across all conditions (CC 0.83–0.85) that this metric is less diagnostic on its own.

\subsection{Physical Interpretability and Articulatory Control}

To validate physical interpretability, we analysed the mapping between generated articulatory trajectories and the acoustic spectrum. Unlike opaque standard TTS, grounding our synthesis in human kinematics enables explicit, interpretable articulatory and accent modifications that are unavailable in opaque systems.

\subsubsection{Evaluation of Articulatory Independence in Minimal Pairs}

To probe physical interpretability, we evaluated the model's ability to distinguish place of articulation while maintaining coarticulatory consistency. Figure \ref{fig:minimal_pair_kinematics} illustrates the generated vertical trajectories for the minimal pair ``pat'' (bilabial stop /p/ onset) and ``cat'' (velar stop /k/ onset). In natural speech, independent articulatory gestures coordinate to produce sound \cite{browman1992articulatory}. Physiologically, /p/ requires active labial constriction with a neutral tongue, whereas /k/ elevates the tongue dorsum toward the velum with minimal labial involvement \cite{ladefoged1996sounds}. The generated traces successfully capture this biological reality. During the onset of ``pat'', the lower lip trajectory ($LL_y$) starts significantly higher to achieve bilabial closure compared to its neutral position in ``cat''. Conversely, for ``cat'', primary motor activity correctly shifts backwards: the tongue dorsum trajectory ($TD_y$) originates at a notably higher elevation for velar closure, positioned well above its resting state seen during ``pat''. Furthermore, the model demonstrates robust physical consistency throughout the shared rhyme sequence (/-æt/). In the shared rhyme (Fig.~\ref{fig:minimal_pair_kinematics}, second row), the jaw ($\text{Jaw}_y$) drops for the open vowel /\ae/ before elevating for the final /t/ closure. Despite minor spatial variations typical of natural speech, these intermediate representations reliably capture expected physiological kinematics.

\subsubsection{Acoustic-Kinematic Realisation of the American Rhotic}

To evaluate the model’s mapping of continuous acoustic signals to complex kinematics, we analysed the generated trajectories for ``cart'' (/k'a:rt/). Because the reference speaker is American, the model is expected to realise the post-vocalic /r/ through a distinct rhotic gesture, universally characterised by the third formant (F3) lowering toward F2 \cite{stevens2001acoustic, delattre1968dialect}.

Figure \ref{fig:cart_rhotic} illustrates this precise acoustic-kinematic mapping. In the acoustic panel, the generated F3 predictably plunges, reaching its maximum lowering around 0.30s. Concurrently, the kinematics exhibit a highly accurate physical response. To execute the retroflex rhotic posture, the tongue tip simultaneously elevates ($TT_y$) and retracts ($TT_x$). This apical gesture is supported by concurrent retraction of the tongue dorsum ($\text{TD}_x$) and body ($\text{TB}_x$), capturing the secondary posterior constriction of the American /r/, while the lowered jaw ($\text{Jaw}_y$) confirms decoupling of tongue shaping from the jaw cycle.

For the final /t/, $F3$ rises as the tongue tip exits the retroflex posture and advances ($TT_x$) toward the alveolar closure, consistent with acoustic generation grounded in coherent kinematics.

\subsubsection{Controllability Using Articulatory Trajectory Modification}
To demonstrate the model's capacity for fine-grained, interpretable control, we performed targeted kinematic manipulations to alter the dialectal realisation of the word ``bird''. The baseline American synthesis exhibits a rhotic gesture during the central vowel, Tongue Tip ($\text{TT}_y$) and Tongue Body ($\text{TB}_y$) elevation acoustically marked by an $F_3$ plunge toward $F_2$, the signature of English rhoticity \cite{delattre1968dialect, stevens2001acoustic}. To synthesise a non-rhotic variant, we flatten the articulatory trajectories across the duration of the vowel tokens, overriding the rhotic gesture and forcing a relaxed, neutral tongue posture. As shown in Figure \ref{fig:bird_modification}, flattening the tongue elevation removes the $F3$ drop in the synthesised audio. $F3$ remains elevated and distinctly separated from $F2$, consistent with standard non-rhotic phonology \cite{ladefoged1996sounds}, demonstrating that the model has learned to associate this gesture with its expected acoustic correlates.

\section{Conclusion}
\label{sec:conclusion}
We presented a controllable source–filter TTS architecture that exposes both sides of speech production as explicit, physically grounded intermediate representations. By conditioning the filter spectrogram on predicted articulatory trajectories supervised via pseudo-targets generated by a multilingual AAI model, we obtain a text-driven synthesis pipeline in which the vocal tract response is anchored in the kinematics of the human articulators. Experiments demonstrate that this physical grounding incurs only a modest spectral-fidelity cost, achieving intelligibility, prosodic fidelity, and naturalness comparable to baselines of similar model capacity. More importantly, the architecture provides quantitative and qualitative evidence of source-filter disentanglement: prosodic manipulation leaves the filter stable, cross-speaker swapping ties pitch to the source speaker and spectral characteristics to the filter speaker, and the predicted trajectories follow known physiological patterns including the American rhotic gesture. We further demonstrate that direct intervention on these trajectories enables interpretable accent modification, converting a rhotic American token into a non-rhotic variant through articulatory trajectory editing alone, with no acoustic post-processing, a form of control unavailable in opaque neural TTS systems. While the proposed architecture trades some raw spectral fidelity for explicit structural disentanglement, this trade-off opens a direction for TTS research in which speech can be generated, manipulated, and analysed through the same physical primitives that govern human speech production. Future work will explore enforcing inter-articulator correlations to respect physical constraints, scaling the model toward state-of-the-art audio quality, and extending the framework to further use cases such as voice conversion and atypical speech.

\vspace{-0.5mm}
\bibliographystyle{IEEEtran}
\bibliography{bibliography}
\end{document}